\pdfoutput=1
\documentclass{JINST}
\usepackage{subcaption}
\usepackage{amssymb}
\usepackage[utf8]{inputenc}
\usepackage[export]{adjustbox}
\usepackage{wrapfig}
\usepackage[font=small,labelfont=bf]{caption}
\usepackage{footmisc} 
\usepackage{tabularx} 
\usepackage[pdftex]{graphics}
\usepackage{comment}
\usepackage{lipsum}
\usepackage{xargs}

\title{Unfolding Neutron Spectrum with Markov Chain Monte Carlo at MIT Research Reactor with He-3 Neutral Current Detectors}

\author{A.~Leder$^{a,g}$, A.~J.~Anderson$^{b,c}$, J.~Billard$^d$, E.~Figueroa-Feliciano$^e$, J.~A.~Formaggio$^a$, C.~Hasselkus$^f$, E.~Newman$^a$, K.~Palladino$^f$, M.~Phuthi$^a$, L.~Winslow$^a$, L.~Zhang$^a$\\
\llap{$^a$}Massachusetts Institute of Technology, 77 Massachusetts Ave. Cambridge, MA 02139, USA\\
\llap{$^b$}Fermi National Accelerator Laboratory, Batavia, IL 60510, USA\\
\llap{$^c$}Kavli Institute for Cosmological Physics, University of Chicago, Chicago, IL, USA 60637\\
\llap{$^d$}Univ Lyon, Université Claude Bernard Lyon 1, CNRS/IN2P3, Institut de Physique nucléaire de Lyon, 4 rue Enrico Fermi, F-69622 France\\
\llap{$^e$}Northwestern University Department of Physics $\&$Astronomy 2145 Sheridan Road, Evanston, IL 60208-3112\\
\llap{$^f$}Dept. of Physics and Astronomy, University of Wisconsin, Madison, WI 53706, USA\\
\llap{$^g$}Corresponding Author\\  

  E-mail: \email{aleder@mit.edu}}
\newcommand{\NCDgas}[0]{$^3_2$He~}
\newcommand{\Cf}[0]{$^{252}_{98}$Cf~}

\abstract
{
   The Ricochet experiment seeks to measure Coherent (neutral-current) Elastic Neutrino-Nucleus Scattering (CE$\nu$NS) using dark-matter-style detectors with sub-keV thresholds placed near a neutrino source, such as the MIT (research) Reactor (MITR), which operates at 5.5 MW generating approximately $2.2\times 10^{18}~\nu$/second in its core. Currently, Ricochet is characterizing the backgrounds at MITR, the main component of which comes in the form of neutrons emitted from the core simultaneous with the neutrino signal. To characterize this background, we wrapped Bonner cylinders around a \NCDgas{} thermal neutron detector, whose data was then unfolded via a Markov Chain Monte Carlo (MCMC) to produce a neutron energy spectrum across several orders of magnitude. We discuss the resulting spectrum and its implications for deploying Ricochet at the MITR site as well as the feasibility of reducing this background level via the addition of polyethylene shielding around the detector setup.  
}
\keywords{Neutron detectors (cold, thermal, fast neutrons); Neutrino detectors; Gaseous detectors;Detector modeling and simulations I }

\begin{document}
\newcommand{\CENNS}[0]{CE$\nu$NS~}


\section{Introduction}
Coherent (neutral-current) Elastic Neutrino-Nucleus Scattering (\CENNS) is a phenomenon whose observation and characterization would probe physics beyond the Standard Model. The recent discovery of \CENNS  by the COHERENT collaboration \cite{Akimoveaao0990} has proven the feasibility of the idea, opening the door to the characterization of \CENNS. \CENNS has recently attracted attention as a gateway to understanding non-standard interactions \cite{Scholberg:2009ha} in the neutrino sector as well as contributing to our understanding of supernova dynamics. Several proposed experiments~\cite{Barbeau,Wong}, including Ricochet~\cite{Anderson:2012}, seek to take advantage of the low-threshold and high-mass detectors already deployed for dark matter direct detection experiments. The Ricochet proposal seeks to utilize a CDMS-style superconducting Zn and/or Ge detector with a target threshold of 100 eV and an initial payload of 1 kg. Zinc superconducting detectors, the subject of ongoing research and development, offer the possibility of intrinsic background rejection due to the difference in quasi-particle interaction dynamics between nuclear and electron recoils~\cite{Billard2016}. In this paper, we explore the possibility of deploying the first phase of the Ricochet experiment 7 meters from the 5.5 MW MIT (research) Reactor (MITR) core. The MITR has an expected core neutrino flux of $2.2 \times 10^{18}$~$\nu$/second corresponding to a \CENNS signal event rate of approximately 1 events/kg/day with low threshold Zn detectors. The proximity to the reactor core comes at the cost of an additional intrinsic background in the form of neutrons, which could mimic a \CENNS signal in Ge/Zn detectors. High-energy reactor neutrons (above 1 MeV) in particular make it through the concrete shielding surrounding the reactor core and can then interact with our detectors. This creates the need for additional polyethylene shielding around the detector to absorb this high-energy neutron background. This underscores the need for accurate information on both the shape and overall normalization of the neutron energy spectrum over a wide range of energies. In order to achieve this neutron monitoring, we deployed a \NCDgas detector that had previously been used to monitor neutrons produced in neutral-current reactions at the Sudbury Neutrino Observatory (SNO) together with incrementally thicker PVC layers in a Bonner cylinder approach. 
    
\begin{wrapfigure}{L}{0.45\textwidth}
\centering
\includegraphics[width=0.45\textwidth]{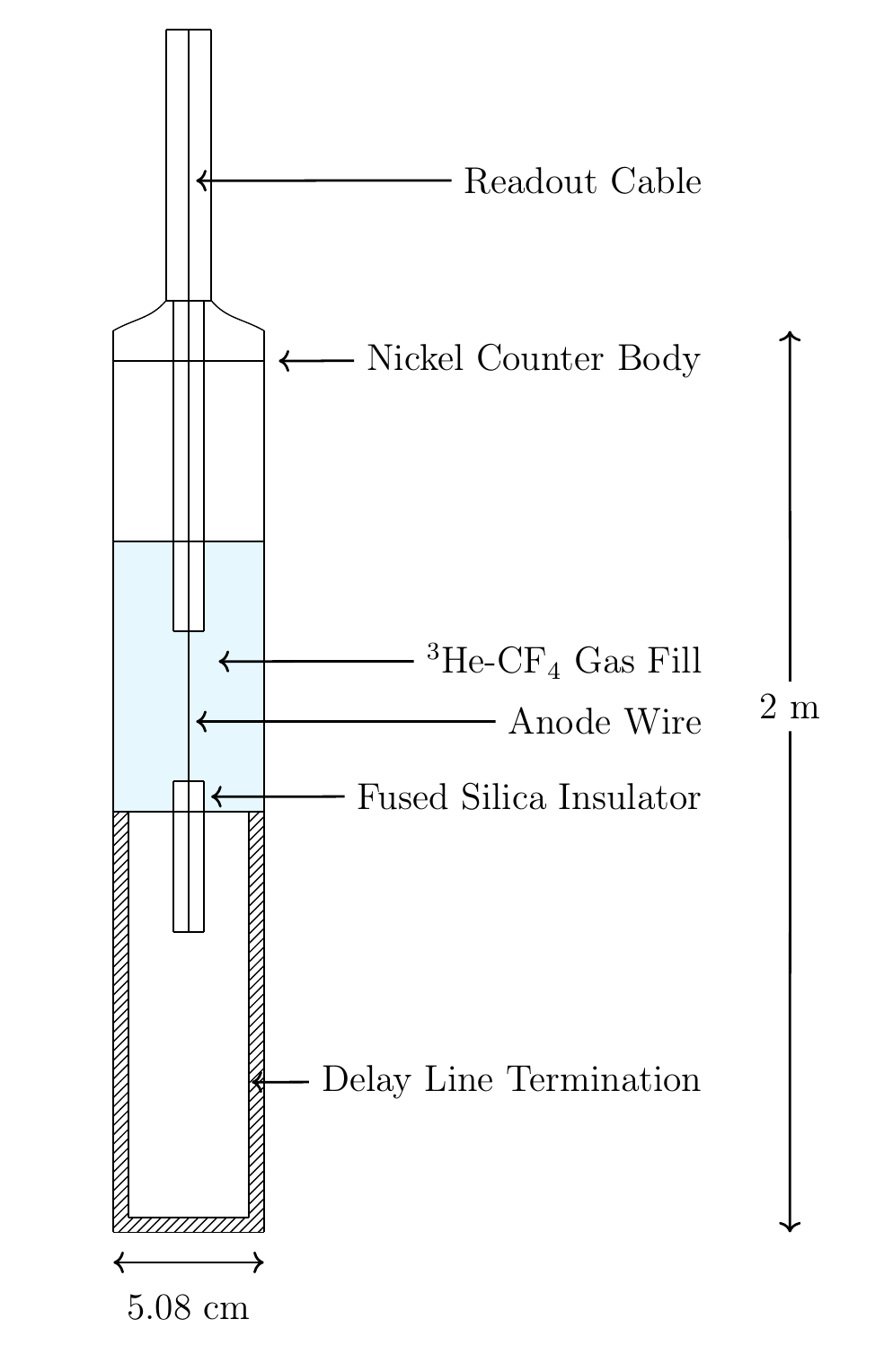}
\caption{Labeled schematic of the NCD detectors taken from \cite{Newmannthesis}}
\label{fig:NCDsetup}
\end{wrapfigure}

    The \NCDgas{} detector as shown in Figure ~\ref{fig:NCDsetup}, has a high neutron sensitivity at thermal neutron energies, but this drops off quickly at higher neutron energies \cite{King}. Akin to a Bonner sphere, by increasing the amount of shielding around the detector one samples progressively higher energy components of the incoming neutron spectrum, which have been thermalized to the sensitive region of the \NCDgas{} detector. For each shielding configuration, we then record energy deposited in the \NCDgas{} detector as well as the rise time of each pulse for use as a pulse-shape discriminator (discussed in Section 2). 
    
After collecting the event rates from the various shielding configurations, we then unfold both the neutron spectral shape and the normalization. This is achieved by creating a series of transfer functions for each shielding configuration in Geant4 by simulating the detector response to various mono-energetic neutron sources as discussed in Section 3. This suite of transfer functions is then used to calculate the likelihood of various binned neutron spectral shapes, which allows us to then optimize the spectral shape and normalization that maximizes the likelihood. For calibration and as a cross-check of our unfolding procedure we also deployed a \Cf source at the MITR site. Finally, we ran additional Geant4 simulations, using our calculated neutron spectrum as an input, which sought to measure the possibility of reducing the overall neutron background to a level below the expected \CENNS signal rate. 

\section{Experimental Setup}\label{sec:experimentalsetup}
\subsection{The Neutral Current Detectors - Physics}
    In order to monitor the neutron flux, we used one Neutral Current Detector (NCD) taken from the SNO experiment ~\cite{Allison:2006ve}, which had previously been used to tag neutrons produced in neutral-current interactions. The NCD consists of one 2-meter-long cylinder filled with a 85:15 percent mixture of \NCDgas{} and CF$_{4}$,\footnote{CF$_{4}$ operates as a quenching gas which stabilizes the avalanche process that occurs when the induced charge gets to within tens of microns of the anode collection wire} with a collection anode wire running coaxially through the gas. The detectors were biased at 1650 V, which represented the highest gain we could use before screening effects from space charge degraded the NCD energy resolution \footnote{Measured to be 44 keV (FWHM) at 764 keV}. Incoming neutrons interact with the \NCDgas{} via the following process:
\begin{equation}
\textrm{n} + ^3_2\textrm{He} \rightarrow \textrm{p} + ^3_1\textrm{H} + \textrm{764 keV}
\end{equation}
	The proton and the triton molecules share the 764 keV$_{kinetic}$ between themselves and then ionize the \NCDgas{}/CF$_{4}$ gas mixture as they lose energy passing through the cylinder. While the ions get collected on the outer grounded shell of the NCD, the resulting induced charge is collected on the anode wire which is then read out using a charge-sensitive pre-amplifier (Canberra Model 2006) fed into a National Instruments ADC (model number: USB-5132). We performed a series of SRIM simulations ~\cite{Ziegler20101818} to determine the proton and triton track length in the \NCDgas{}, determining an average track length of 7.5 and 2.5 mm respectively ~\cite{Beltran:2011sha}. We tested 6 different shielding configurations corresponding to an overall shielding thickness that ranged from 0 up to 7.23 cm radially. For each of the 6 shielding configurations we collected data with the reactor on/off and with a \Cf neutron calibration source. 
 

\subsection{The Neutral Current Detectors - Event Topology and Selection}
\begin{figure}
  \centering
  \begin{subfigure}[b]{0.45\textwidth}
    \includegraphics[width=\textwidth]{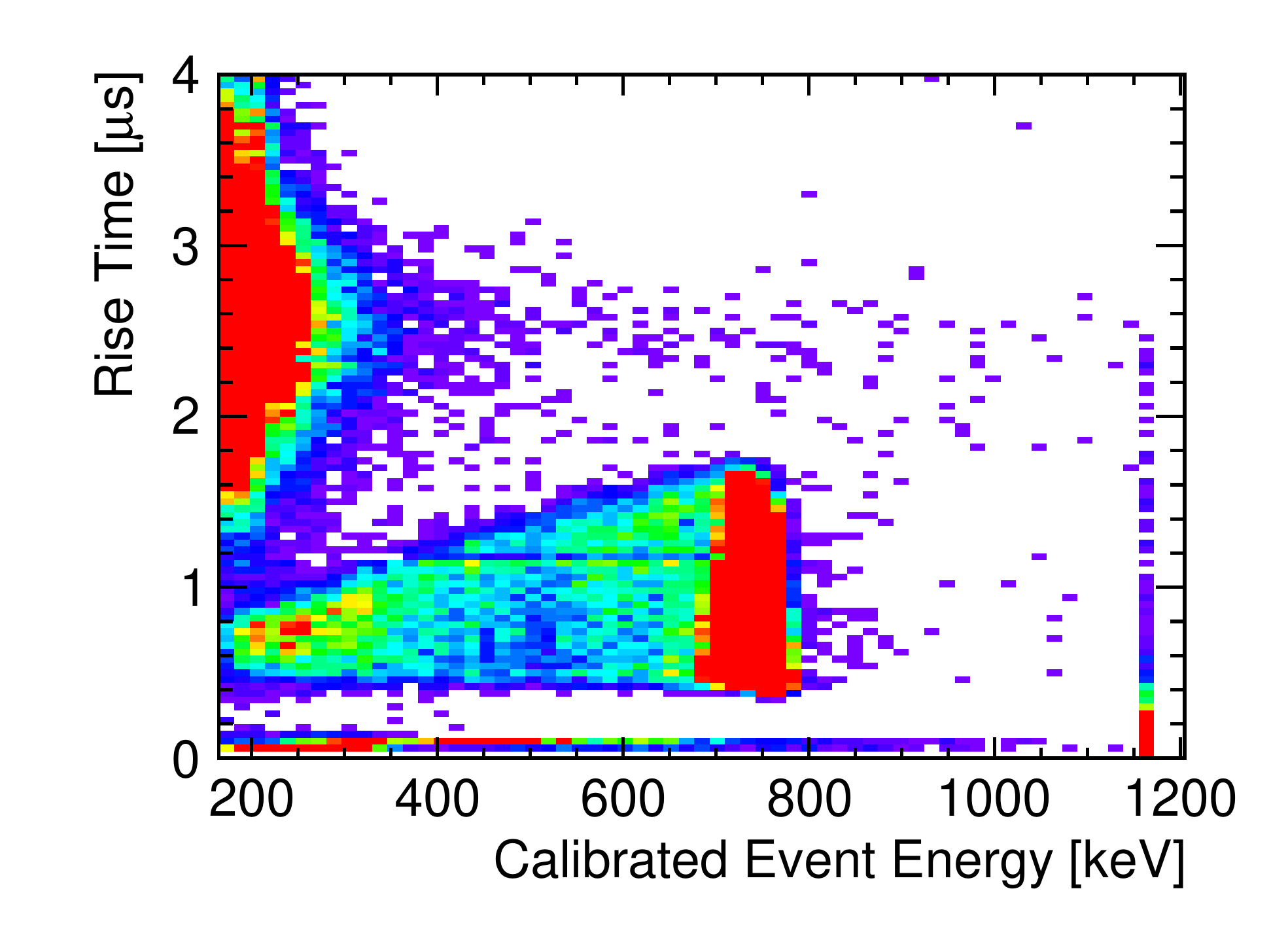} 
    \caption{The trapezoid plot showcasing the four event populations present and their relative densities in our detector setup. The cut on Neutron Capture (NC) events was defined by a series of 4 linear cuts over the NC region, with particular care taken on the region between the LIE and NC event regions}  
    \label{fig:triangleplot}
   \end{subfigure}
   \begin{subfigure}[b]{0.45\textwidth}
     \includegraphics[width=\textwidth]{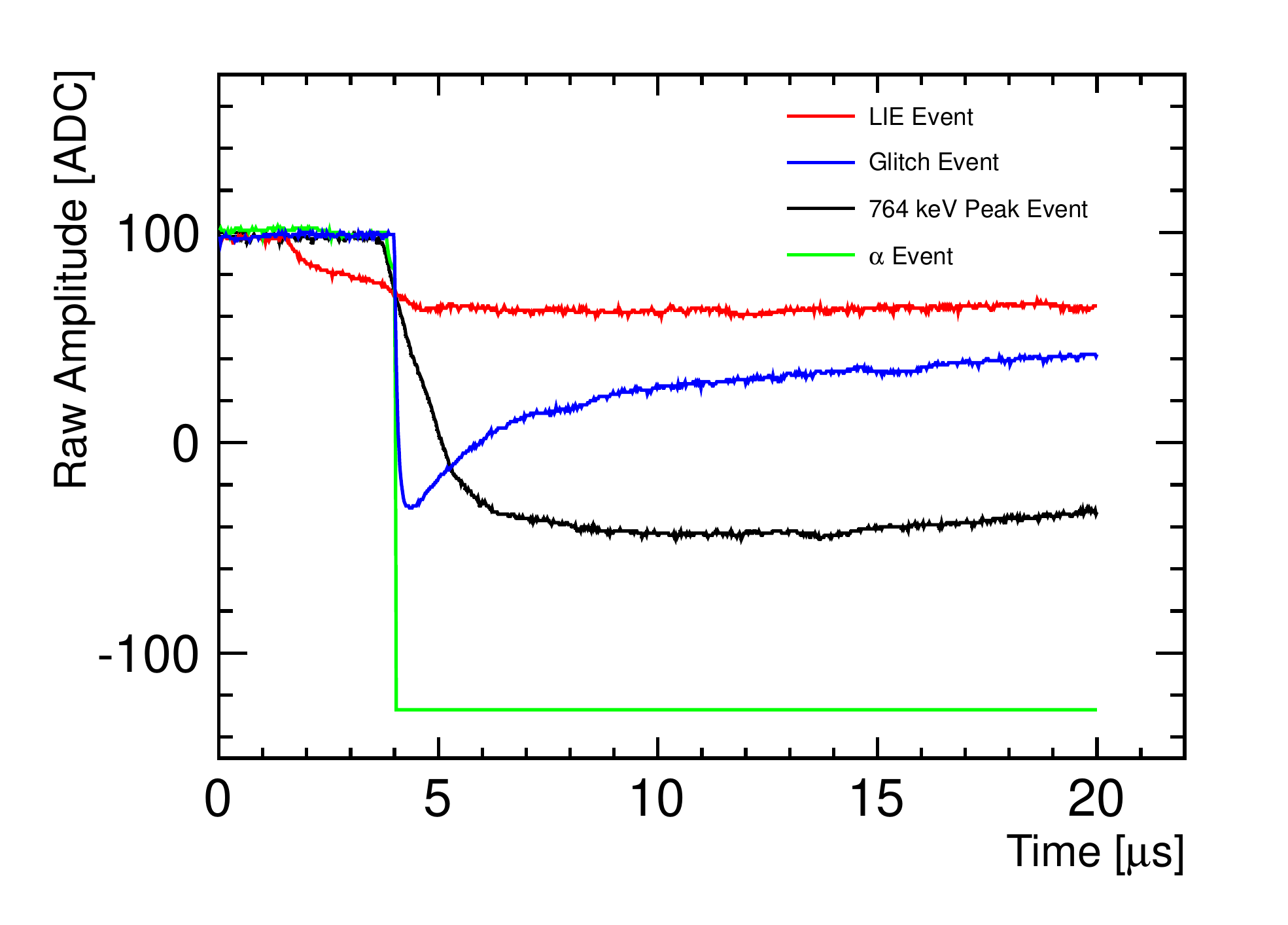} 
     \caption{Each of the example pulses shown here was drawn from one of the four event regions/topologies visible in figure ~\ref{fig:triangleplot}. The large difference in rise times stem from differences in the orientation of event tracks while the difference in energies for NC events stem from the capture of one or both of the \NCDgas{}-Neutron reaction products} 
    \end{subfigure}
    \caption{The lower trapezoid corresponds to the signal neutron captures while the band on the left correspond to Low Ionizing Events (LIE). The band of events seen at approximately 1200 keV corresponds to high energy alpha energy depositions saturating the readout electronics. Events with extremely short rise times form a population of glitch events which were discarded as background in this analysis}
\end{figure}    

When an event interacts with the NCDs and ionizes the \NCDgas{}/~CF$_{4}$ mixture, the $dE/dx_{event}$ plus the track orientation determines the event time length, while the overall event amplitude is determined by the total energy of the proton/triton pair collected by the anode wire.

\begin{figure}
  \centering
  \begin{subfigure}[b]{0.45\textwidth}
    \includegraphics[width=\textwidth]{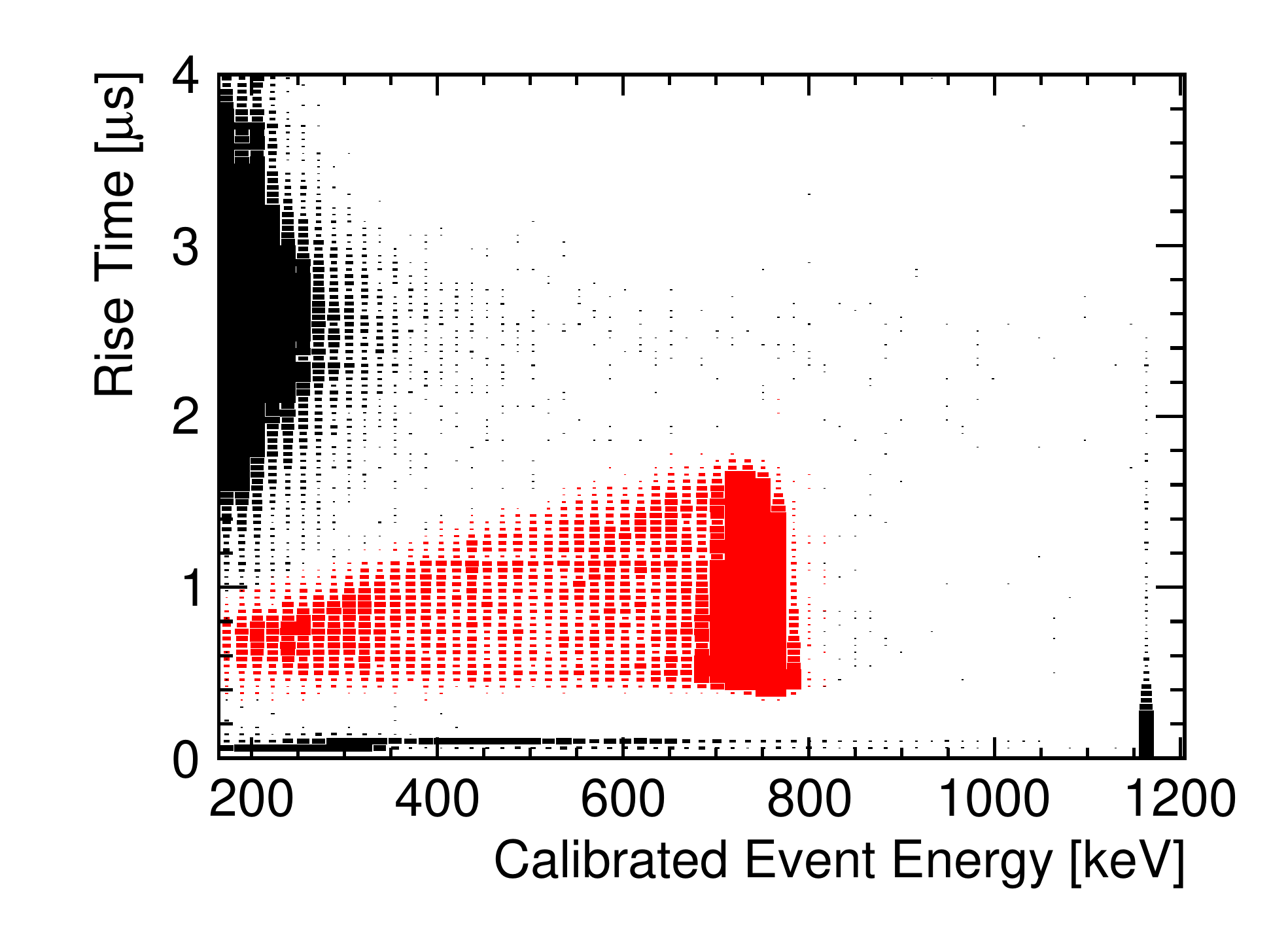} 
    \caption{Rise-time versus Energy plot taken from a typical NCD run. Red events indicate those events that passed the data quality cuts that identify these events as neutron induced (NR) events, while all black events represent background LIE/Glitch/$\alpha$ events.  
    \label{fig:EnergyTimewithcuts}}
  \end{subfigure} 
  \begin{subfigure}[b]{0.45\textwidth}
    \includegraphics[width=\textwidth]{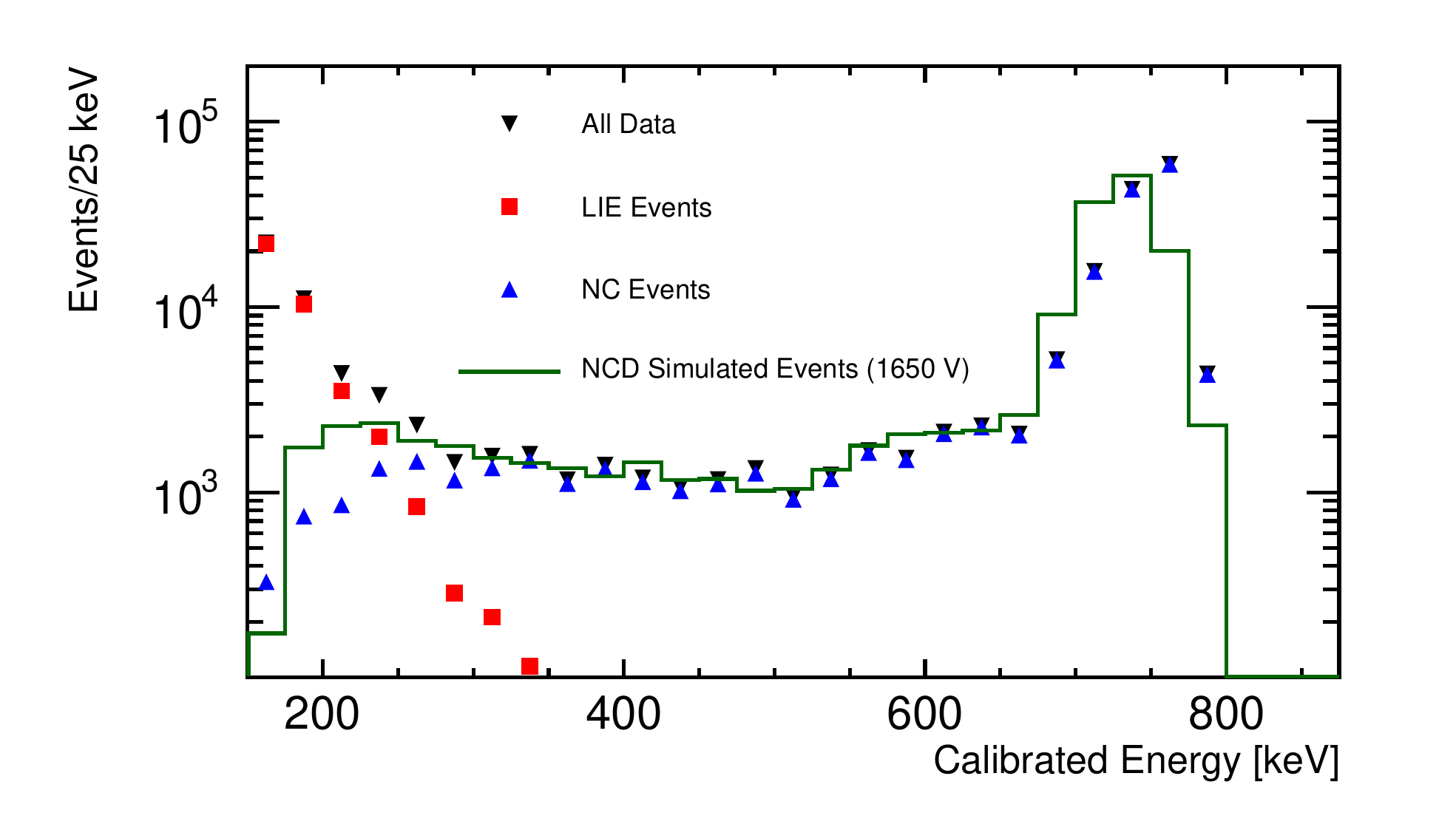}
    \caption{Calibrated spectrum of events in the neutron capture region. The 764 peaks together with the 564 and 191 keV shoulders were fitted to calibrate the NCD response. The 764 keV peak was fitted using an Exponentially Modified Gaussian (EMG) \cite{Grushka}, while the 564 and 191 keV shoulders were fitted with error functions. Note that the error bars on the NR/LIE spectrum may be too small to see}
     \label{fig:RoughSpectrum}

  \end{subfigure}    
  \caption{NCD nuclear capture event criteria and calibrated NCD spectra (after selecting nuclear capture events) as well as a simulated NCD spectrum. In both plots one can see the 764, 573 and 191 keV features that are the hallmarks of NCD NR events}
  \label{fig:NCDoutput}
\end{figure}

From each event trace, two variables were estimated: the rise time\footnote{We defined the rise time as the time it takes for the event trace to go from 10 to 70 percent of the maximum amplitude} and the event amplitude. From these data, we could then produce a "trapezoid plot" (as seen in Figure ~\ref{fig:triangleplot}) with two clear populations of events visible: with fast, high-energy events defining neutron captures and generally slower lower energy background events. The main source of background events for the NCDs stem from Compton scattered light particles such as electrons and muons with longer rise times coupled with lower event amplitudes allowing for their classification as Low-Ionizing Events (LIE). We also observed two additional background signal populations with extremely short rise times, corresponding to micro-discharges/glitch events and events that saturated the DAQ system at 1.2 MeV corresponding to $\alpha$ deposition events. For a given energy of event, one can also note the range of rise times which stems from the particular orientation of the event track, with perfectly parallel events resulting in very short rise times and perpendicular events having the longest rise times. Lower energy neutron events produce a smaller range of rise times because of incomplete event collection due to the walls of the NCDs. 
The NCDs were calibrated by using a three-point calibration fitting scheme using the 191 keV, 564 keV and 764 keV event peaks/shoulders which correspond to: full energy of just the proton collected, full energy of just triton collected and full reconstruction of the entire event-with both proton and triton induced ionization collected on anode wire respectively.
   The event-selection criteria were defined by a trapezoidal region defined by a series of four linear boundaries as shown in Figure ~\ref{fig:EnergyTimewithcuts} in this rise-time/energy-deposition space. The four linear boundaries were defined by looking at the data collected using the maximum PVC thickness as these data had the clearest separation between LIE and neutron events due to the thick shielding around the NCDs. 
   
\begin{figure}
\centering
\includegraphics[width=0.80\textwidth]{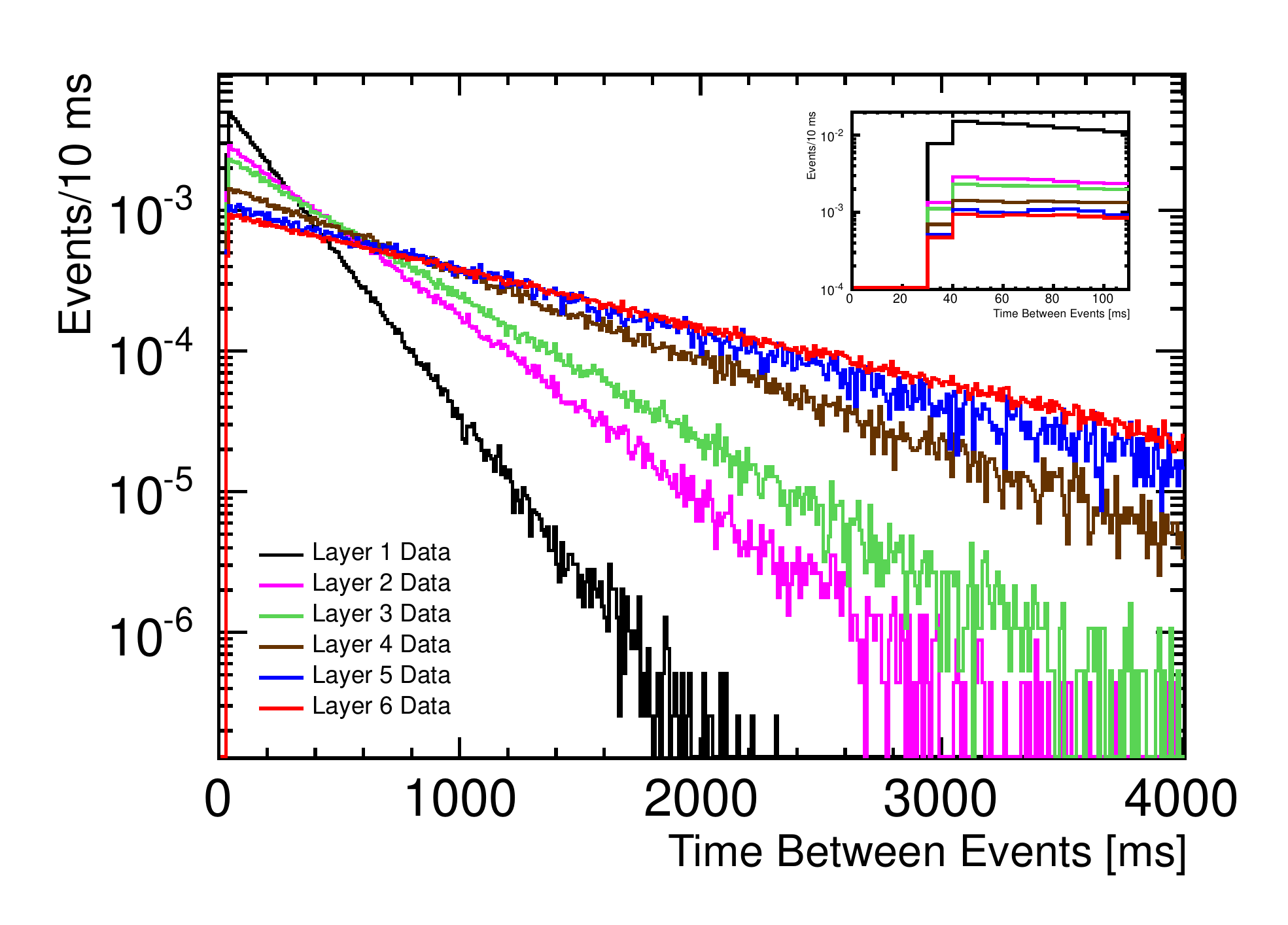}
\caption{Time between subsequent nuclear recoil events for all the shielding configurations we ran, normalized to 1. One can clearly see the monotonic decrease in event rate as a function of layers, consistent with a strong thermal component to the neutron spectrum. $\textbf{Inset}$: Same plot focusing on the short time between events to highlight the 30 ms dead time in our detector setup}
\label{fig:TimeBetween}
\end{figure}

The gain and shape of the trapezoid region was stable from run to run and over the course of 24+ hour long runs, which justified our use of the same selection criteria for all the data sets collected. The rounded edge of the nuclear recoil band at 764 keV visible on Figure~\ref{fig:RoughSpectrum} was due to space charge screening effects. The space charge effects were simulated in GEANT4 using the framework established by ~\cite{Tseungthesis}. At lower anode voltages, space charge effects are minimized as evidenced by the sharp cutoff in the NCD spectrum at 764 keV (visible in Figure~\ref{fig:NCDoutput}), while at lower event energies the space charge effect is negligible. Reasonable agreement was achieved between the simulated and the collected NCD spectrum even at higher anode voltages. Space charge effects were determined to only have a negligible effect on the systematics on the overall event rates extracted from the NCDs due to the excellent signal/background ratio at 764 keV. In order to calculate the event rate accounting for dead time and small changes in the event rate over time, the time between events was measured and fit to an exponential distribution. From Figure~\ref{fig:TimeBetween}, one can see that the data is well-described by an exponential distribution and that the detectors had a dead time of approximately 30 ms. These extracted event rates were then fed into the deconvolution procedure outlined below.  Background NCD rate measurements, which included cosmic-ray induced neutrons,  with the reactor off were subtracted from the reactor on NCD data. In order to determine the effective cosmic ray overburden at the MITR for a possible future Ricochet deployment, we collected cosmic ray background data using desktop muon counters ~\cite{Axani2016} during a 24 hour period where the reactor was turned off. We determined the effective overburden to be 1.5 m.w.e. 

\subsection{Geant4 simulation}
Given measurements of the reactor neutron rate with various numbers of PVC layers surrounding the detector, we are interested in recovering both the shape and normalization of the neutron spectrum which requires unfolding techniques. Although there are various approaches advocated in the literature, nearly all require the calculation of a \emph{transfer function} $T^{j}(E_i)$, which relates the incident neutron flux $\phi$ to the event rate $R$ measured by the NCD. In our case, for each shielding configuration $j$, we require a transfer function, $T^{j}$, defined over the same interval as the binned energy spectrum $E_i$, in order to estimate the theoretical event rate:
\begin{equation}
R_{theory}^{j} \simeq \sum_{i=1} \int_{E_i}^{E_{i+1}} \phi(E_i) T^{j}(E, n)~dE.
\end{equation}

We assume that the neutron flux is flat in lethargy space, \footnote{Lethargy is the logarithmic ratio of energy before/after a collision, given by the relation \begin{equation} u = ln(\frac{E_0}{E}) \end{equation}} while showing an inverse energy dependence in energy space throughout a bin. The transfer functions were calculated over a neutron energy range encompassing thermal neutrons up to high energy reactor neutrons. The high-energy cutoff (10~MeV) for the reconstruction bins was selected because less than 1 percent of the emitted neutrons have energies greater than 10~MeV for a thermal reactor core spectrum centered at 1.4~MeV. We then use maximum likelihood to fit the neutron energy spectrum in energy bins to the observed rates for each shielding configuration as discussed further in Section 4. 

We estimate the transfer function of the detector and shielding setups using a Monte Carlo simulation based on Geant4 10.00.p02~\cite{Allison:2006ve}. To simulate the hadronic physics, including neutron interactions, we use a modular physics list including the QGSP\_BERT\_HP model containing the ''high-precision'' neutron physics simulation that uses version 4.4 of the G4NDL data. In addition, our physics list incorporates the ``G4ScreenedNuclearRecoil'' process that models screened electromagnetic nuclear elastic scattering and is important for accurately simulating the propagation of the proton and triton after a neutron capture on $^3$He~\cite{Mendenhall2005420}. Unlike the default models in the QGSP\_BERT\_HP physics list, the G4ScreenedNuclearRecoil processes give physically realistic results for ion track lengths and energy loss in the detector gas, which were also found to be in reasonable agreement with results from the widely-used SRIM software~\cite{Ziegler20101818}.

The simulation includes the layers of PVC shielding as described in the section~\ref{sec:experimentalsetup}, correctly accounting for the geometry of the PVC pipes when resting on the ground. Although it is not needed for the transfer function analysis, the simulation also implements models of the charge propagation, space-charge effects, and the preamplifier electronics, similar to~\cite{Beltran:2011sha}.
To simulate the transfer functions needed for the unfolding analysis, we simulate 10 million monoenergetic neutrons with each of the 6 shielding configurations and at 34 logarithmically-spaced energies between $10^{-8}$~MeV and $10^2$~MeV. The neutrons in the simulation are generated on a cylindrical surface immediately surrounding the outermost layer of shielding, with a direction that is chosen isotropically from the surface of the shielding directed inward toward the detector. We have also determined that bias in our calculations due to non-concentric compared to perfectly concentric alignment of the various shells is small due to the approximate radial symmetry of the NCD setup. Another effect of this radial symmetry is that the NCD setup is insensitive to the initial direction of the reactor background neutrons. In addition, the data collected by the NCDs pointed to a strong thermalized component to our neutron spectrum at the MITR, which supports the assumption that our neutron flux will be closer to isotropic.

\begin{figure}
\centering
\includegraphics[width=0.75\textwidth]{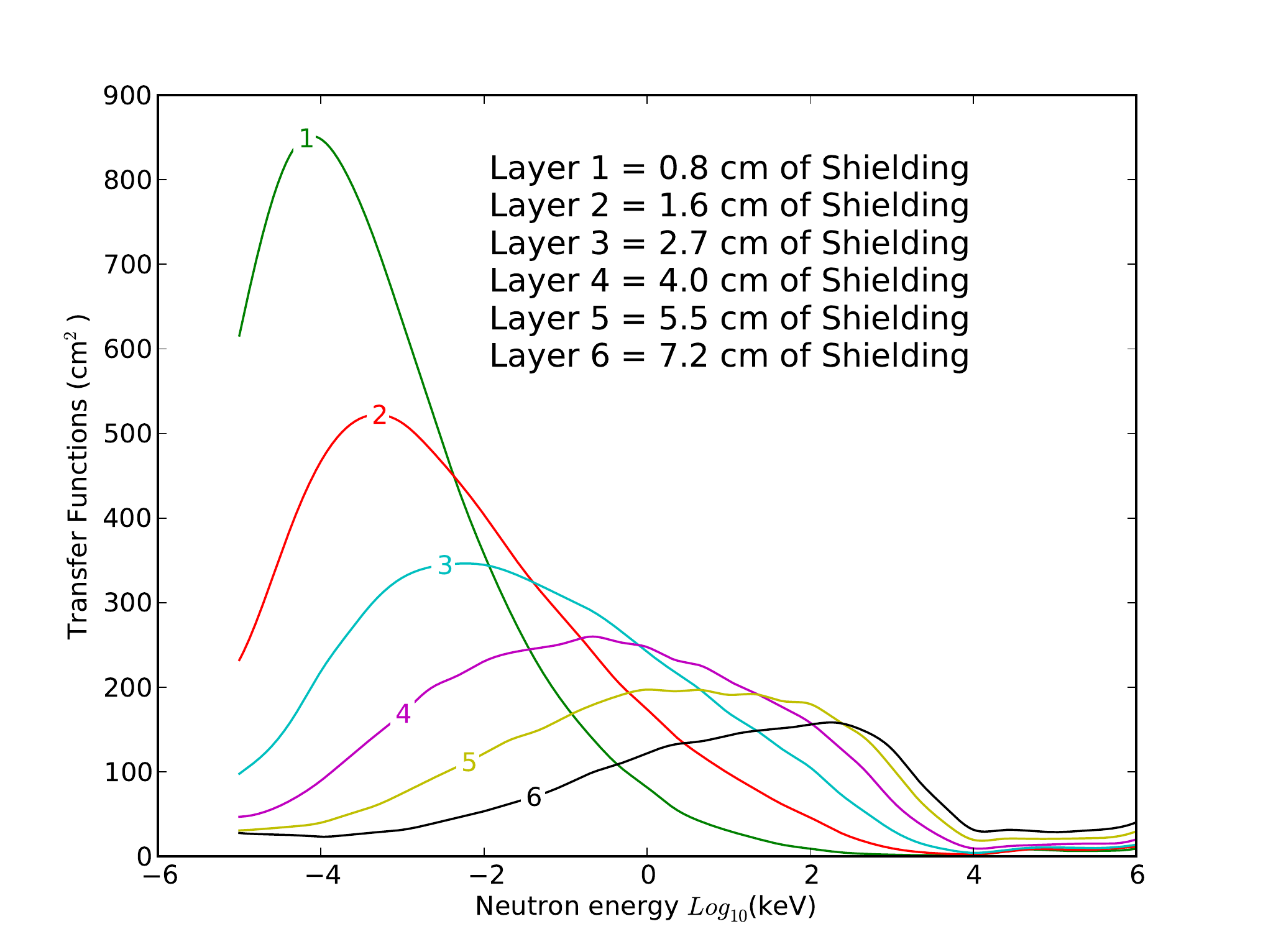}
\caption{Transfer functions for each shielding configuration as a function of incident simulated monoenergetic neutron energy - the large degree of overlap between these transfer functions prompted our use of an MCMC to deconvolve the data. The transfer functions are normalized to the surface area of the shielding configuration used to calculate them.  
\label{fig:TransferFunctions}}
\end{figure} 

Figure~\ref{fig:TransferFunctions} shows the transfer functions for each layer of shielding as a function of energy. From these plots one can note a few important features, namely that we only gain significant sensitivity for fast "core" neutrons with 5 or 6 layers of shielding. In addition, we note a significant degeneracy between the transfer functions for each shielding configuration, which limits the sensitivity of the unfolding analysis, which motivated the idea of using a MCMC analysis as discussed below.

\subsection{Likelihood analysis}

\begin{figure}
\centering
\includegraphics[width=0.70\textwidth]{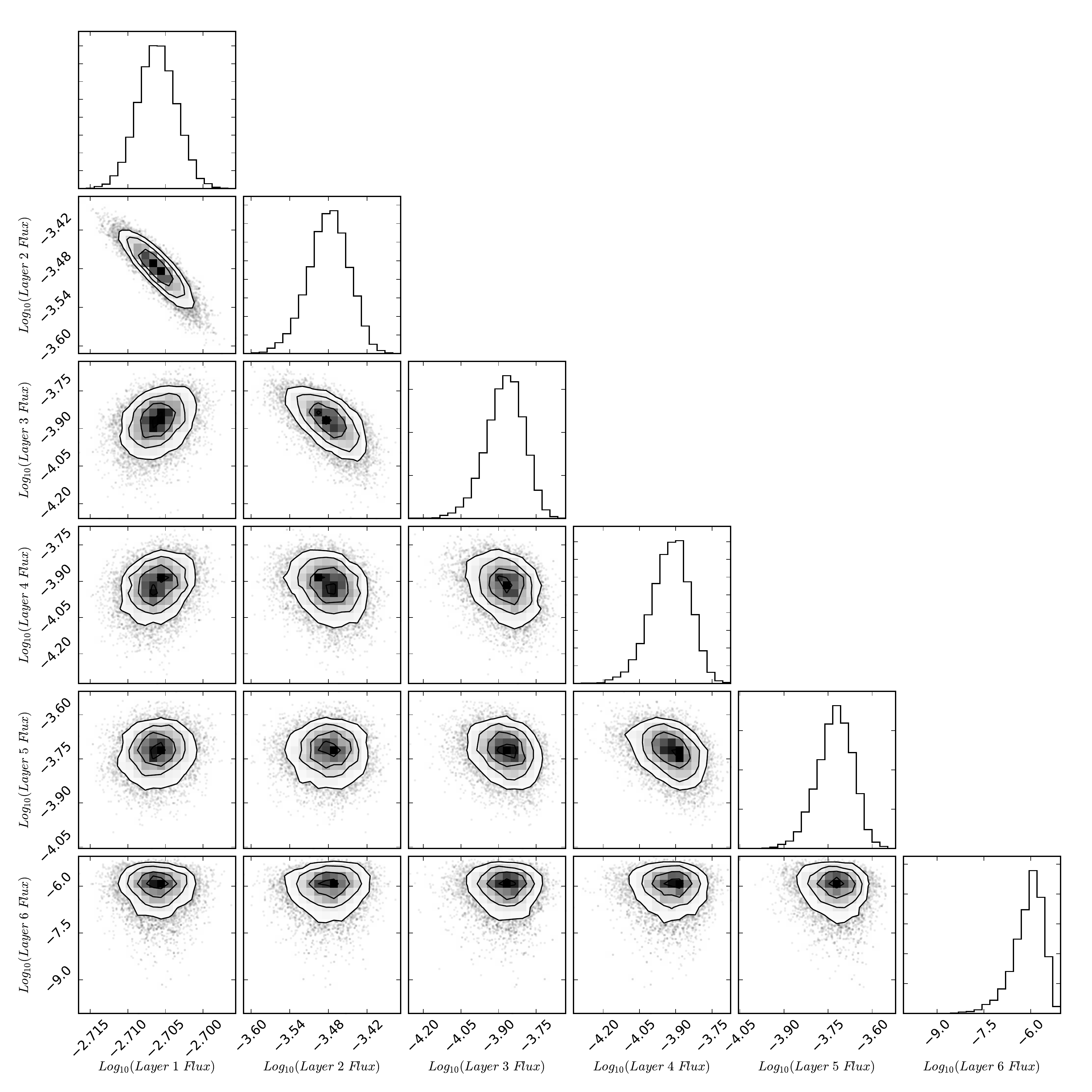}
\caption{Correlation plots for accepted 6-bin MCMC points. Note that the values of each of the MCMC points are shown on a log scale here and in particular bins 1 and 2 have strongest correlations to the other, consistent with the large overlap in the transfer functions \label{fig:MCMCcorrelations}}
\end{figure}

Using the transfer functions (see Figure~\ref{fig:TransferFunctions}) calculated from the Geant4 simulations in the previous section, we can calculate an expected theoretical event rate for each shielding configuration, given a binned neutron spectrum shape. By minimizing the difference between the theoretical and observed event rates we can then extract a binned neutron spectrum. Due to the high correlation observed between the bins (see Figure~\ref{fig:MCMCcorrelations}) and to ensure that the best fit $\chi^{2}_{fit}$ that we found truly represents a global best fit we use the maximum entropy method. 

\begin{figure}
\centering
\includegraphics[width=0.75\textwidth]{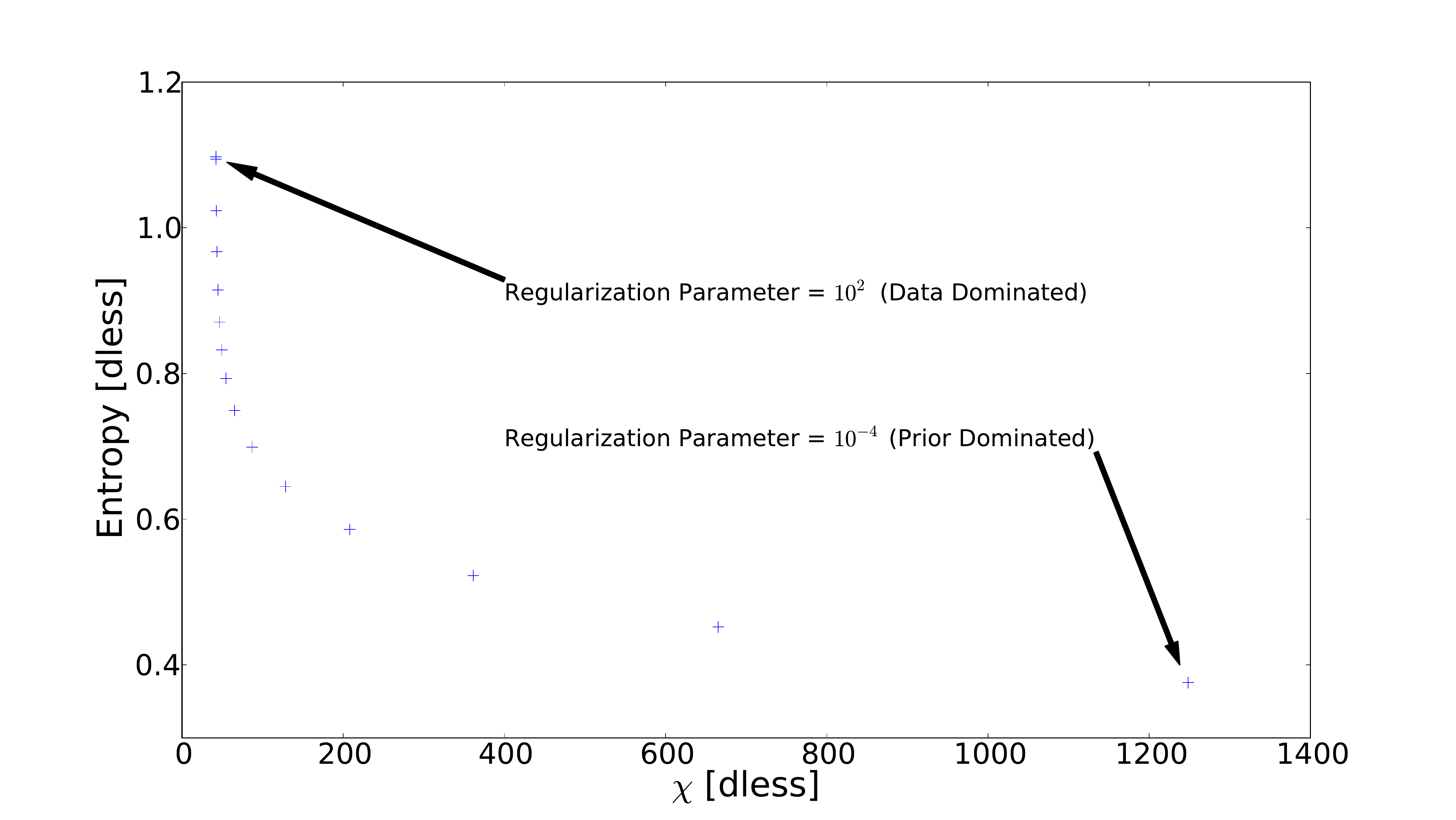}
\caption{$\chi$-Entropy Plot used to calculate the optimal value of the regularization parameter. The optimal value for the regularization parameter was determined to be $10^{-3}$ by examining the point with the largest curvature. The regularization parameter controls whether the fitting procedure we are using is data or prior dominated. 
\label{fig:ChiEntropyplot}}
\end{figure}

First, we define a likelihood $\mathcal{L}$ function for the theoretical event rates for a given binned spectrum ($\phi_{i}$) via the following relations: 

\begin{equation}
\chi^{2}_{fit} = \sum_{i=1}^{n}
{\frac{(R^{obs}_{i}-R^{theory}_{i}[\phi (\vec{E})])^2}
{\sigma^{2}_{i}}}
\end{equation}
\begin{equation}
S = - \sum_{i=1}^{m}p_{i}\log(p_i) \quad\textrm{with}\quad p_i = \frac{\phi_{i}(E)}{\sum_{i}\phi_{i}(E)}
\end{equation}
\begin{equation}
\mathcal{L}_{\phi} = \exp{\left(-\frac{\chi^{2}_{fit}}{2}+\frac{S}{2\textit{w}}\right)}
\label{eq:2p6}
\end{equation}
where $i$ corresponds to the particular NCD shielding configuration and $S$ corresponds to the entropy. In this context, $S$ represents another quantity that needs to be simultaneously minimized by maximizing the likelihood function. In Equation ~\ref{eq:2p6}, $\textit{w}$ corresponds to a regulation parameter which determines if our extracted spectrum is prior- (small $\textit{w}$) or data-dominated (large $\textit{w}$). A prior-dominated deconvolution would result in essential features smoothed out of the unfolded spectrum as the unfolding algorithm would converge to the prior~\footnote{In this analysis, our prior is the result of the Minuit fitting (maximum entropy) plus our choice of regularization parameter}. On the other hand, a data-dominated unfolding can result in large unphysical fluctuations in the resulting unfolded spectrum. We wish to find a value for $\textit{w}$ that minimizes the resulting $\chi^{2}_{fit}$ with the additional condition that small changes in $\chi^{2}_{fit}$ only result in minimal changes to S. In practice, this means finding the point on the $\chi-S$ curve (see Figure ~\ref{fig:ChiEntropyplot}) with the largest curvature. It should be noted that this prescription only gives us an order-of-magnitude estimate for the best value for $\textit{w}$. By varying the regularization parameter along this curve we were also able to calculate the systematic error due to this method, which is included in the quoted systematic error in Table ~\ref{tab:diff_rate}

We assume Poisson fluctuations in the measured values of the $R^{obs}_{i}$, which allows us to approximate the likelihood as a multivariate Gaussian. We also assume that the neutron flux follows an exponential spectral shape within a bin (flat in lethargy space). By finding the binned spectral shape that minimizes $\mathcal{L}$ via Minuit~\cite{Verkerke}, we generate a preliminary binned neutron spectrum; however, due to the high degree of overlap between the various transfer functions this results in highly-correlated spectral neutron bins, and additional minimization is needed to find a global best fit. In order to better characterize this strong correlation between neutron bins and determine the distribution of expected neutron background events at the MITR, we employed a Markov Chain Monte Carlo (MCMC) analysis based on the framework described in Billard et al.~\cite{Billard2011}.

\section{MITR flux reconstruction}
\subsection{Measurements undertaken at the MITR}

\begin{figure}
\centering
\includegraphics[width=0.75\textwidth]{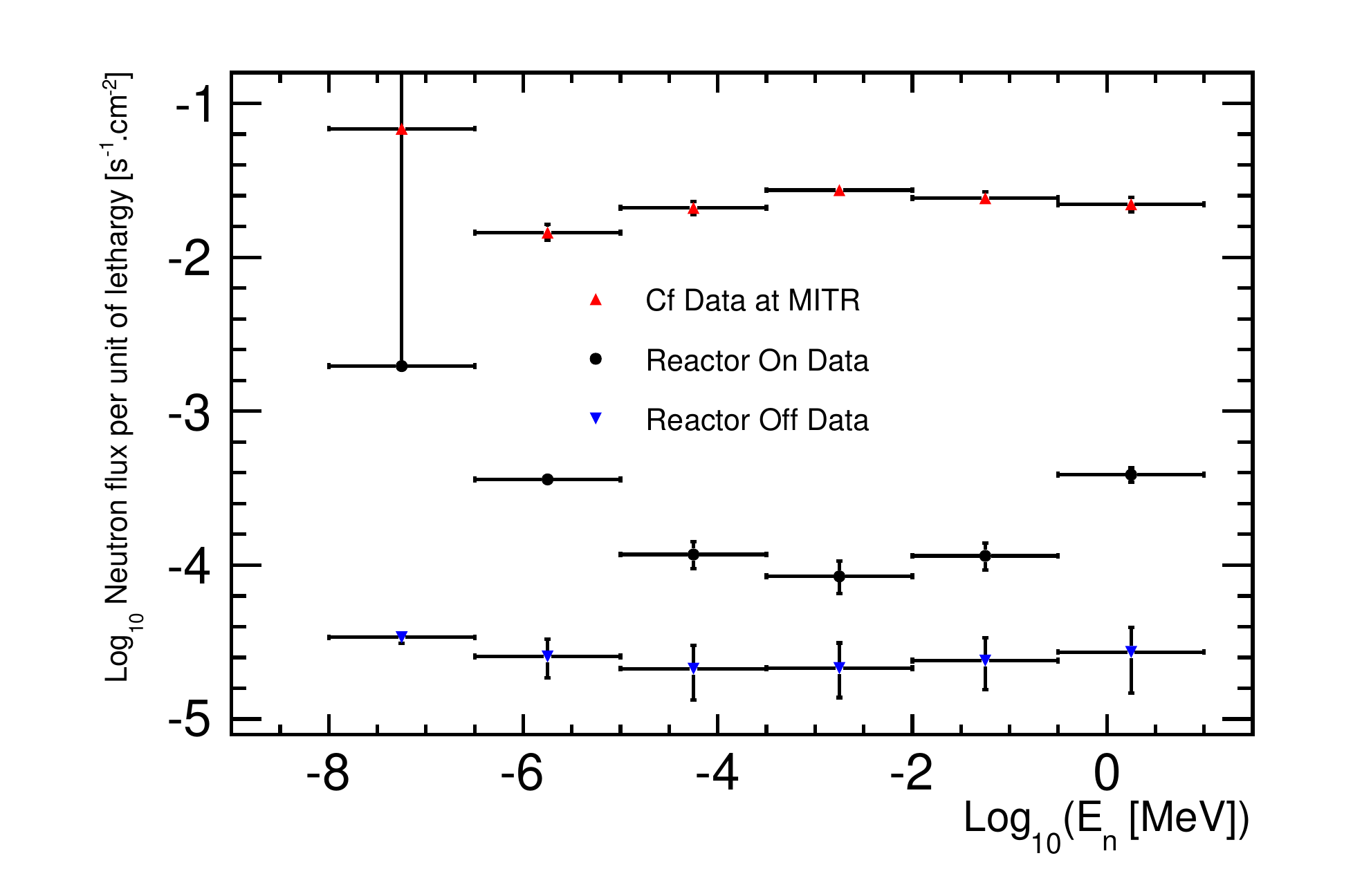}
\caption{Deconvolved spectra using the NCD data collected with reactor on, off and with a \Cf neutron source deployed at the MITR. The vertical error bars here correspond to the a $\pm$ 1 sigma confidence interval based on the MCMC simulations of the reconstructed flux}
\label{fig:Finalspectra}
\end{figure}

Over the course of May-October 2015, we undertook three sets of measurements at the MITR reactor corresponding to the reactor on/off together with a \Cf calibration neutron source. Using the unfolding techniques discussed above we were able to generate the following neutron spectra - see Figure ~\ref{fig:Finalspectra}. 
In particular, for the \Cf calibration data we noticed the peak position shifted downward in energy, which we attribute to thermalization of the neutrons coming from the source. It is also important to note that strong thermal component inherent in all the unfolded spectra. This suggests that the surrounding concrete/shielding converts the raw neutron flux into an isotropic neutron gas; however, for all spectra there still exists a strong high energy (> 1 MeV) component to the neutron spectrum, and is it these high energy neutrons that then will contribute the most to any neutron background for a possible Ricochet deployment at the MITR.

\subsection{Neutron Event Rate for Ricochet Deployment}
As discussed in $\cite{Billard2016}$, Zn superconducting detectors offer the ability to distinguish between nuclear and electromagnetic recoils due to the differences in quasi-particle propagation inside the crystal lattice. Throughout these calculations, we foresee deploying a 1 kg Zn detector with a 100 eV threshold, which represents an achievable target threshold set by the EDELWEISS experiment~\cite{EDELWEISSCollaboration2016}. Based on the calculations in~\cite{Anderson:2012}, we determine the expected \CENNS event rate in Zn at the MITR facility to be approximately 1~events/kg/day at 7~meters from the core. Using the spectra extracted from the NCD data we performed another series of Geant4 simulations with a 1 kg Zn payload deployed in a Adiabatic Demagnetization Refrigerator (ADR) configuration. We simulate neutrons isotropically at a radius of 64 cm from the center of the ADR, drawing from an exponential distribution in energy space across each of the bins used in the MCMC deconvolution. These primary events are then passed through a Geant4 model of the ADR setup to record the energy depositions of any neutrons. By then examining the number of events that fall in the Region of Interest (ROI) between 100 eV$_{NR}$ and 1 keV$_{NR}$ we then weigh each event by its corresponding flux value from the MCMC fit to then extract an expected neutron background event rate. The high neutron background event rate with no shielding highlights the need to include shielding around our detector setup for any deployment at the MITR. This prompted us to perform an additional simulation surrounding the ADR with 30 cm, both radially and on the top/bottom, of bare and boron-doped (5 $\%$ by mass) polythene shielding to determine the feasibility of reducing the neutron background rate to the level of the expected \CENNS event rate. Systematic errors for this shielding simulation were calculated by comparing the neutron stopping power for 0 to 30 cm of poly shielding between 400 keV and 10 MeV to the reduction in flux as seen in \cite{TOMASELLO201070}. Our systematic error indicates that the calculated neutron event rate represents more of an upper limit on the actual neutron event rate that we could expect for a Ricochet deployment at MITR. With 30 cm of polyethylene shielding around the detector, our calculated background neutron event rate becomes comparable to the \CENNS event rate, with the actual neutron background event rate being possibly even lower. By taking advantage of MITR duty cycle, coupled with knowing the precise power levels over the course of 1 to 5 years, we have shown it is still possible to extract a \CENNS discovery signal even when the signal to background ratio is as high as 1:5 (see Table 3 in ~\cite{Billard2016}). With additional shielding, our neutron background rate could be brought down still more, improving our \CENNS sensitivity further. 
\begin{table}[ht]
\centering
\resizebox{\textwidth}{!}
{\begin{tabular}{ |c|c|ccc| }
\hline
Rates (per kg per day) & \CENNS Rate & MITR On & MITR Off & Cf \\
\hline
0 cm poly shielding & 1.0 &$36.3^{+3.2 (stat) + 6.8 (sys)}_{-3.5 (stat) -26.0 (sys)}$ & $5.0^{+1.3}_{-1.7}$ & $(4.5^{+.2}_{-.3}) \times 10^3$ \\
\hline
30 cm poly shielding & 1.0  & $3.2^{+0.4 (stat) +0.5 (sys)}_{-0.3 (stat) -2.1(sys) }$ & --- & --- \\
\hline
30 cm borated shielding & 1.0  & $2.4^{+0.4 (stat) +0.4 (sys)}_{-0.2 (stat) -1.3(sys) }$ & --- & --- \\
\hline
90 \% CL (0 cm borated shielding) & 1.0  & < 41 & < 7  & < 5 $\times 10^3$ \\
\hline
90 \% CL (30 cm borated shielding) & 1.0  & < 2.9  & --- & --- \\
\hline
90 \% CL (30 cm bare poly shielding) & 1.0 & < 3.8 & --- & --- \\
\hline
\end{tabular}}
\caption{Summary of the neutron background rate expected by a Ricochet deployment at the MITR for the three experimental configurations tested in this paper. All borated tests conducted with 5$\%$ (by mass) boron-doped poly shielding}
\label{tab:diff_rate}
\end{table}

\section{Conclusion}
\begin{figure}
\centering
\includegraphics[width=0.75\textwidth]{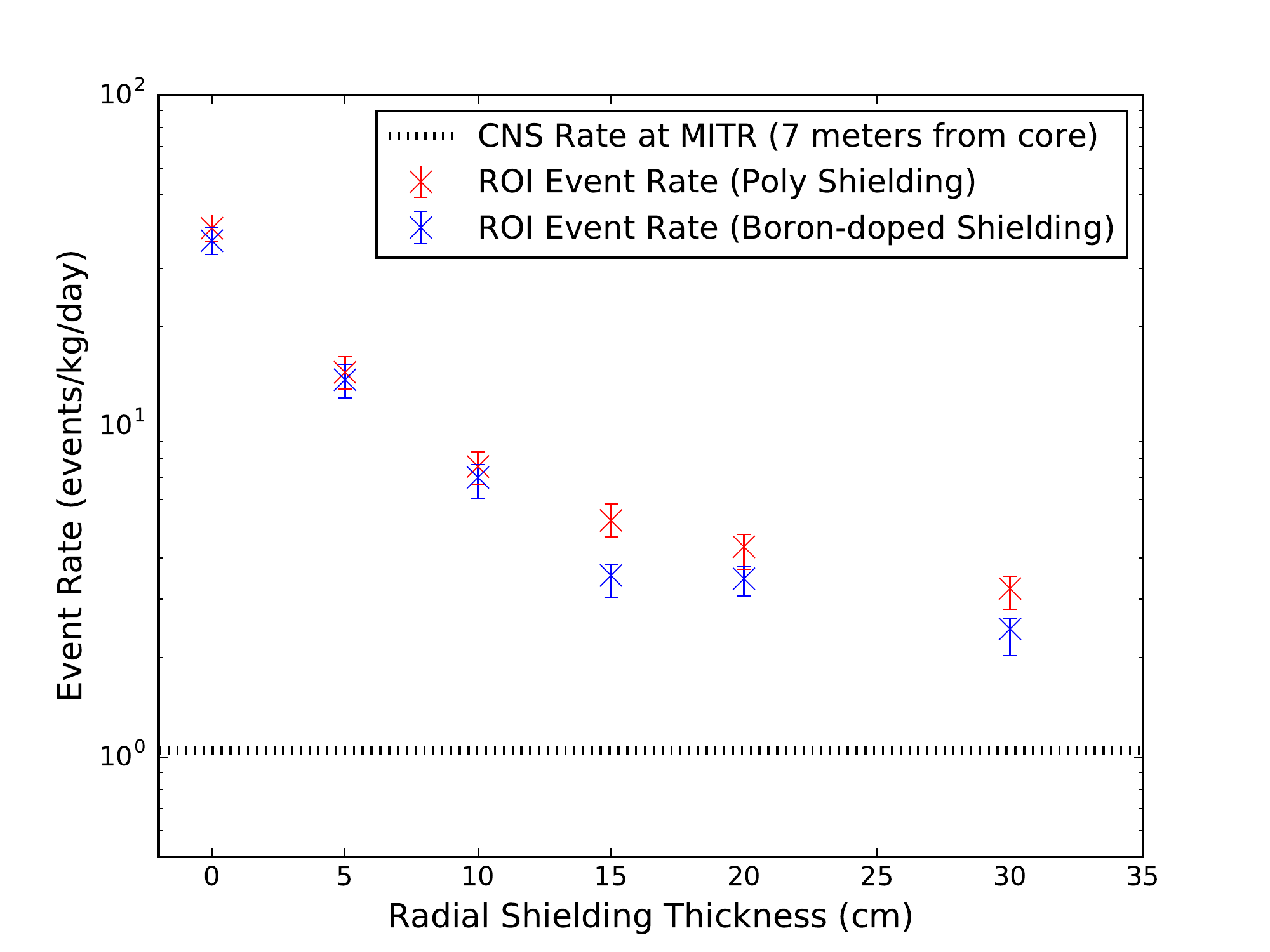}
\caption{Event rate in the Region of Interest (between 0.1 and 1 keV) as a function of various thicknesses of Poly shielding (Normal and Boron-doped) around the sensitive detector region. Note: the error bars on this plot only correspond to statistical errors}
\label{fig:shieldingrates}
\end{figure}

We successfully deployed NCDs at the MITR site in order to measure the neutron background across several orders of magnitude to lay the groundwork for a possible future stage 1 deployment of the Ricochet experiment. After applying pulse shape discrimination event-selection criteria we extracted an event rate for each of the 6 different NCD Bonner cylinder shielding configurations, each of which were used to probe different components of the overall neutron spectrum. We applied an optimized maximum likelihood coupled with a MCMC analysis framework to characterize the high correlations between the various neutron bins in the final reactor on, reactor off and \Cf calibration data. Using the final unfolded neutron spectrum we simulated a 1 kg Ricochet deployment with a Zn target at the MITR with and without 30 cm of polyethylene shielding. While the raw unfolded spectrum points to an intrinsic neutron background rate higher than the expected \CENNS signal rate. With the addition of 30 cm of shielding this background event rate was brought down close to the \CENNS signal level as shown in Figure ~\ref{fig:shieldingrates}. While the low signal to background ratio makes a \CENNS signal detection search challenging, we have shown in \cite{Billard2016}, that it is possible to extract a \CENNS discovery signal even when the signal to background rates are close to 1 to 5. With a similar analysis, coupled with additional passive shielding, the MITR represents not just a good location for continued Ricochet Zn Bolometer R$\&$D testing, but also a potential additional \CENNS signal site.

\acknowledgments{}
We wish to thank all the personnel, in particular Thomas Bork and Dane Kouttron, of the MITR complex for all their help in setting up and running our experiment. We also wish to thank the SNO collaboration and in particular Hamish Robertson for lending us the NCD detectors. This material is based upon work supported by the U.S. Department of Energy, Office of Science, Office of Nuclear Physics under Award Numbers FG02-97ER41041 and DE-FG02-06ER-41420.

\newpage

\bibliography{references}

\end{document}